\documentclass[conference]{IEEEtran}
\usepackage{cite}
\usepackage{amsmath,amssymb,amsfonts}
\usepackage{algorithmic}
\usepackage{graphicx}
\usepackage{textcomp}
\usepackage{xcolor}
\def\BibTeX{{\rm B\kern-.05em{\sc i\kern-.025em b}\kern-.08em
    T\kern-.1667em\lower.7ex\hbox{E}\kern-.125emX}}
    
\usepackage{graphicx}
\usepackage{enumitem}
\usepackage{booktabs, makecell, multirow}
\usepackage{subcaption}
\usepackage[ruled,vlined]{algorithm2e}
\usepackage{float}
\usepackage{bbm}
\usepackage{amsthm}
\usepackage{balance}

\newcommand{\pelican}{\textit{Pelican}\xspace}
\newcolumntype{P}[1]{>{\centering\arraybackslash}p{#1}}
\begin{document}

%\title{Privacy Implications of Personalized Machine Learning Models in Context-Aware Mobility Applications}
%\title{Privacy Preserving Machine Learning Model Personalization for Distributed Mobile Services}
\title{Preserving Privacy in Personalized Models for Distributed Mobile Services}

% \author{\IEEEauthorblockN{Under Submission at ICDCS 2021}}
\author{\IEEEauthorblockN{Akanksha Atrey}
\IEEEauthorblockA{
\textit{University of Massachusetts Amherst}\\
aatrey@cs.umass.edu}
\and
\IEEEauthorblockN{Prashant Shenoy}
\IEEEauthorblockA{
\textit{University of Massachusetts Amherst}\\
shenoy@cs.umass.edu}
\and
\IEEEauthorblockN{David Jensen}
\IEEEauthorblockA{
\textit{University of Massachusetts Amherst}\\
jensen@cs.umass.edu}
}

\maketitle

\begin{abstract}
% The ubiquity of mobile computing has led to the proliferation of mobile devices. Modern mobile applications and services are context-aware providing customized content to users based on their present context (e.g., location). Context-aware applications thrive on the ability to predict future contexts to pre-fetch content or make recommendations. With the advancement of hardware edge technologies, it is common for such mobile applications to employ machine learning to process, reason and make tailored predictions by learning personalized models. Despite the advantages of personalization, the level to which privacy guarantees hold in personalized models is still unknown as most prior work has only considered privacy-preserving training and inference for multi-user domains.
% In this work, we argue that personalized mobility prediction models can leak privacy since they encode mobility behavior unique to the user. We demonstrate several attribute inference-based privacy attacks on personalized time-series ML models that enable sensitive information to be learned. Our results show that personalized models can leak privacy at inference time with a higher leakage at smaller spatial scales. We thus provide a robust and simple approach based on scaling the sensitivity to outputs to mitigate the privacy risks of personalized models at inference time in mobile applications. We show the efficacy of the solution via empirical evaluation.
The ubiquity of mobile devices has led to the proliferation of mobile services that provide personalized and context-aware content to their users. Modern mobile services are distributed between end-devices, such as smartphones, and remote servers that reside in the cloud. Such services thrive on their ability to predict future contexts to pre-fetch content or make context-specific recommendations. An increasingly common method to predict future contexts, such as location, is via machine learning (ML) models. 
%To do so, they increasingly use machine learning models to predict future contexts such as location and make context-specific recommendation.
Recent work in context prediction has focused on ML model personalization where a personalized model is learned for each individual user in order to tailor predictions or recommendations to a user's mobile behavior.  
%The state-of-the-art in machine learning involves model personalization where a general model is tailored ("personalized") to each individual user in order to tailor predictions or recommendations to a user's mobile behavior. 
While the use of personalized models increases efficacy of the mobile service, we argue that it increases privacy risk since a personalized model encodes contextual behavior unique to each user. To demonstrate these privacy risks, we present several attribute inference-based privacy attacks and show that such attacks can leak privacy with up to 78\% efficacy for top-3 predictions. We present \pelican, a privacy-preserving personalization system for context-aware mobile services that leverages both device and cloud resources to personalize ML models while minimizing the risk of privacy leakage for users. We evaluate \pelican using real world traces for location-aware mobile services and show that \pelican can substantially reduce privacy leakage by up to 75\%.
\end{abstract}

\begin{IEEEkeywords}
%privacy, mobile devices, mobile applications, personalized modeling, mobility, machine learning
cloud-based mobile services,  personalized ML models, privacy, deep learning, context-awareness
\end{IEEEkeywords}

%Section 1
\section{Introduction}

The ubiquitous nature of smartphones and smart devices, such as wearables, have led to a plethora of online mobile services in various domains including fitness, entertainment, news and smart homes. Such mobile services tend to be distributed between the end-device and the cloud with front-end components running on the devices as mobile applications and back-end components running on cloud servers. Modern mobile services are often context-aware to provide tailored content or service to users based on their current context. For example, it is common for a restaurant recommendation service to use location as its context when recommending nearby eateries. While the use of {\em current} context in mobile services is common, mobile services have begun to use machine learning (ML) models to predict {\em future} contexts (e.g., a user's next or future location(s)) and provide tailored recommendation based on these prediction (e.g., suggest directions or store closing time of predicted future location).

Machine learning has been used in mobile services for tasks such as next location prediction \cite{zhao2020go}, medical disease detection \cite{dai2019machine} and language modeling \cite{yoon2017efficient}. The popularity of deep learning has established the use of aggregated data from a large number of users to train and deploy a general ML model that makes predictions for context-aware services for a broad range of users. A more recent trend in the field is to use personalized models on a per-user basis rather than a general model to further improve the efficacy of the service. In this scenario, rather than using a single ML model for all users, a model is personalized for each user using training data specific to the user. For instance, a user's frequently visited locations in a mobile service or a user's viewing history in a streaming service can be used to develop personalized ML models. %Our approach for model personalization uses transfer learning methods that utilize the inductive biases of a multi-user ML model and tailor it to a distinct user using limited single-user context traces.

While model personalization is a growing trend in mobile and Internet of Things services, in this paper, we examine the implications of such an approach on the privacy of individuals. We argue that personalized ML models encode sensitive information in the single-user context traces used as training data and mobile services that use such personalized models can leak privacy information through a class of privacy attacks known as model inversion. Model inversion attacks exploit a trained ML model to infer sensitive attributes \cite{fredrikson2015model}. While ML researchers have studied inversion attacks in other contexts, they have not been studied or demonstrated for time-series models that are commonplace in mobile applications. Our work formalizes and demonstrates such attacks for personalized mobile services by showing how they can leak sensitive context (i.e. location) information about a user. To the best of our knowledge, privacy implications of personalized models in distributed mobile services have not been previously studied.

Motivated by the need to ensure the privacy of personalized ML models, we present \pelican, an end-to-end system for training and deploying personalized ML models for context-aware mobile services. Our system enhances user privacy by performing sensitive personalized training on a user's device and adding privacy enhancements to personalized models to further reduce and prevent inversion attacks from leaking sensitive user information. Our system is also designed to allow low overhead model updates to improve model accuracy while safeguarding privacy. Finally, our system leverages the device and cloud architecture of mobile services when personalizing ML models to enhance user privacy. In design and implementation of \pelican, we make the following contributions:

\begin{enumerate}[label=\textbf{C\arabic*}, topsep=2pt]
    \item We adapt low-resource transfer learning methods to train and execute personalized ML models on resource-constrained mobile devices for mobility applications. Our approach utilizes the inductive biases of a multi-user ML model and tailors it to a distinct user using their limited context traces. 
    %We leverage general models trained on the cloud to gather high-level patterns and then learn personalized patterns using single-user data streams. 
    Our work draws inspiration from existing work on transfer learning-based personalization of language models \cite{yoon2017efficient}.
    \item We formalize practical inference-based privacy attacks on personalized models using model inversion \cite{fredrikson2015model}. We consider ways in which an adversary can reconstruct private historical information using only trained personalized mobility prediction models. Our work formalizes model inversion attacks for time-series based ML models with application in the mobility domain.
    \item We quantify the efficacy of these privacy attacks on mobile services that use personalized models. Our findings demonstrate that such attacks can leak private historical mobility patterns with up to 78\% accuracy for top-3 predictions. We find that the leakage is higher for smaller spatial scales and independent of user mobility behavior.
    \item We present the design of \pelican, an end-to-end privacy preserving personalization framework. We propose a robust enhancement to mitigate inference-based privacy attacks based on scaling the output probability distribution at inference time. %We demonstrate how this enhancement is simple yet effective in thwarting against inference-based attacks through empirical analysis. 
    We empirically evaluate \pelican on low-level and high-level spatial mobility scales using a campus dataset and show that \pelican is able to reduce privacy leakage up to 75\%.
\end{enumerate}

% The rest of the report is organized as follows. Section \ref{sec:background} discusses the background pertaining to mobile services, personalized ML models and privacy. In Section \ref{sec:pmandprivacy}, we present our approach for building personalized models for mobility prediction and introduce inference-based privacy attacks on time-series based personalized models. We present the privacy attack in Section 5. In section 6, we propose a solution to thwart inference-based attacks and demonstrate its efficacy through empirical analysis. We conclude the report in Section 7.

%Section 2
\section{Background}
\label{sec:background}
In this section, we present background on context-aware mobile services and the use of ML models in such services.

\textbf{Context-Aware Mobile Services.}
Our work considers mobile services whose service components are distributed across mobile devices and a back-end cloud. It is typical for mobile services to be context-aware and tailor the service based on current or future contexts.
%Advances in mobile computing have led to the proliferation of smart mobile devices with approximately 76\% of users in advanced economies employing them \cite{silver2019pew}. Smart mobile devices are used in various ways that are unique to the user to offload or assist human thinking \cite{mosa2012systematic, pejovic2014interruptme, barr2015brain}. 
In recent years, context-aware mobility applications, such as location-based social networking and ride-sharing applications, have gained popularity.
Context can be defined as any information used to characterize interactions with the environment or situation of an entity %\cite{dey2001understanding , wang2004context} 
and can be broadly categorized into temporal, spatial and social. 
%Temporal contexts utilize temporal information of the environment. Halvey et al. demonstrate the importance of time-based contexts in modeling users' mobile behavior \cite{halvey2006time}. Spatial contexts span over geographic settings in which events occur and social contexts refer to how users react to their physical environment.
A common type of context-aware service utilizes the user's current or future location to offer location-aware mobile services. Unless specified otherwise, our work assumes location to be the primary context used by the distributed mobile service.

% \begin{figure}
%     \centering
%     \includegraphics[width=0.6\columnwidth]{figures/distributed_service.png}
%     \caption{Caption}
%     \label{fig:my_label}
% \end{figure}

\textbf{Mobility Prediction.}
%what is it -- use history of recurring locations to predict next.
In addition to using current context such as location, many services now use next location prediction techniques to predict future location(s) that a user will visit and offer recommendations based on future contexts. For instance, a mapping service may predict commute times to the next location a user is predicted to visit. Next location prediction techniques capture the spatial and temporal correlations between human mobility patterns. 
%With the growing popularity of mobile computing, many location-aware mobile applications are using next location prediction. The primary goal behind mobility prediction is to capture the spatial and temporal correlations between human mobility patterns. 
Since humans tend to follow particular routines and habits, learning their mobility behaviors can assist many domains from recommendation systems to urban design. Human mobility can be defined through a series of location and time-varying features. Consider a set of features $x_t = \{l, e, d\}$ with location $l$, entry time $e$ and duration $d$ at time $t$. The mobility prediction problem can be defined as follows: given a set of previous sequences $s_u = \{x_1, x_2,...x_t\}$ for user $u$, estimate location $l_{t+1}$ of user $u$ at the next time step.

\textbf{Time-Series ML for Next Location Prediction.}
%background and example on next location prediction
Prior work in next location prediction has focused on using variants of Markov models, Hidden-Markov models and tree-based classification models to learn the sequential nature of mobility \cite{gambs2012next, mathew2012predicting}. 
%These models are able to learn the sequential nature of mobility but can get computationally expensive with growing data or dependencies. 
With the emerging capabilities in deep learning to handle temporal or spatial input, recurrent neural networks (RNN) have been proposed for mobility prediction \cite{song2016deeptransport}. %, du2016recurrent, liu2016predicting, lin2017deep, feng2018deepmove, liao2018predicting, jiang2018deep}. 
RNNs have the ability to capture sequential data where each sample is dependent on its previous samples. More recently, a variant of RNNs, long short term memory (LSTM) models \cite{hochreiter1997long} have shown state-of-the-art performance in predicting human mobility \cite{kong2018hst, zhao2020go, trivedi2020empirical, feng2020pmf}. Unlike RNNs, LSTMs have the ability to learn and remember long-term dependencies in the data. Deep learning-based models generally employ mobility trajectories of many users to learn generic human mobility patterns and are capable of handling large prediction spaces typical of general mobility models. 

\textbf{Model Personalization.}
%what is a general model? what is a personalized model? why should we personalize for mobile app (to tailor service to each user)
A common approach for using ML models in mobile services (e.g., for predicting future contexts) is to train a general ML model using aggregated training data from a larger number of users. Such a model encodes behavior of a large group of users and can predict the future behavior of a user who resembles one in the training set. A recent trend, however, is to employ a personalized model that is designed for a specific user over the use of a general model. Personalized models can encode specific behavior exhibited by an individual user and offer better efficacy over an aggregated model.
%Humans employ applications and services in ways that are unique to themselves. A common characteristic among such applications is their ability to collect and communicate continuously streaming data. This provides a path for building models and systems that are personalized for the user (i.e. user-centric). Personalized models are preferred for their ability to tailor content for individual users, privacy-preserving nature as such models are not prone to differential privacy-based attacks, and computational efficiency. 
In recent years, machine learning methods for personalization have been proposed in various domains including autonomous vehicles \cite{vallon2017machine}, health \cite{rudovic2018personalized}, %, khademi2019personalized}
%fitness \cite{dijkhuis2018personalized, zhou2018personalizing}
and natural language processing \cite{yoon2017efficient}. 
Recently, Sarker et al. explored the effectiveness of ML models for predicting personalized context-aware smartphone usage \cite{sarker2019effectiveness}. They evaluate numerous ML algorithms and find that tree-based models, such as random forests, are the most effective for building personalized context-aware models.
Personalized modeling in mobility has been generally conducted via Markov models \cite{gambs2012next}. %chen2014nlpmm, do2014and}. 
More recently, Feng et al. developed personal adaptors for personalized modeling with LSTMs \cite{feng2020pmf}.
%Our work adapts deep learning-based transfer learning to learn personalized models for context-aware mobility applications. To the best of our knowledge, transfer learning on deep models to achieve personalized models have not been employed in context-aware mobility applications.

\textbf{Machine Learning Privacy.}
%what is a inversion attack more generally? what is an inversion attack for next location prediction?
Machine learning models are vulnerable to privacy attacks and our work argues that model personalization increases privacy risks for users.
Two of the primary privacy attacks in ML are \textit{membership inference attacks} \cite{shokri2017membership} and \textit{attribute inference attacks} \cite{fredrikson2015model}. Membership inference attacks aim at inferring whether a data sample was present in the training set. Given a model $M$ and some data point $x$, the goal is to infer whether $M$ used $x$ during training. This attack is particularly problematic when using sensitive data sets. For instance, if a ML model is trained on a cancer data set and an adversary is able to infer whether a user was in this data, it will reveal the user's health status. In this work, we focus on attribute inference attacks, namely a model inversion attack. Model inversion attacks aim at inferring sensitive features using a trained model. Given a model $M$ and some features $\{x_2, x_3...,x_n\}$, the goal is to reconstruct the value of some sensitive feature $x_1$. This is problematic when the data set contains sensitive features such as the location of a user.
Model inversion attacks were first proposed by Fredrikson et al. \cite{fredrikson2014privacy} to exploit linear regression pharmacogenetic models to infer patient genotype. There have been various subsequent papers on understanding the broader risk of such attacks \cite{fredrikson2015model, wu2016methodology, hidano2017model, yang2019neural, he2019model, zhang2020secret}. %\cite{fredrikson2015model, wu2016methodology, hidano2017model, veale2018algorithms, basu2019membership, yang2019neural, he2019model, wang2019beyond, zhang2020secret}. 
Wu et al. proposed a game-based formalization of inversion attacks for any ML model yet claimed that privacy leakage from such attacks is context dependent \cite{wu2016methodology}. Our work formalizes model inversion attacks for time-series applications with a focus on mobility. We focus on reconstructing users' historical mobility patterns using a next location prediction model.

%Section 3
\section{Personalized Models and Privacy Implications}
\label{sec:pmandprivacy}
In this section, we first describe our approach for model personalization based on transfer learning of deep learning models and then describe our model inversion privacy attack on such personalized models.
%In this section, we discuss the traditional approach for mobility prediction and present two approaches for personalization. We focus on personalized modeling for next location prediction. 

\subsection{ML-driven Next Location Prediction}
Predicting the next location(s) based on historical locations is a fundamental mobility task that is useful in a broad range of mobile services. We describe three approaches based on deep learning to perform next location prediction. 
%A typical assumption in location prediction is access to historical \textit{trajectories} (temporally extended sequences of locations) of the user. Location-aware mobile applications collect such data which can be easily transformed into trajectories and related variables, such as time of entry to location and duration of stay at the location. 

\begin{figure}
\centering
    \begin{subfigure}[b]{0.5\columnwidth}
    \includegraphics[width=\columnwidth]{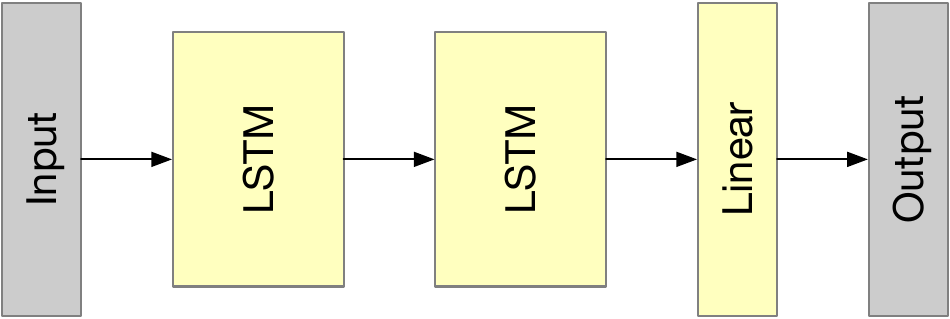}
    \caption{}
    \label{fig:traditional_arch}
    \end{subfigure}
    \hfill
    \centering
    \begin{subfigure}[b]{0.65\columnwidth}
    \includegraphics[width=\columnwidth]{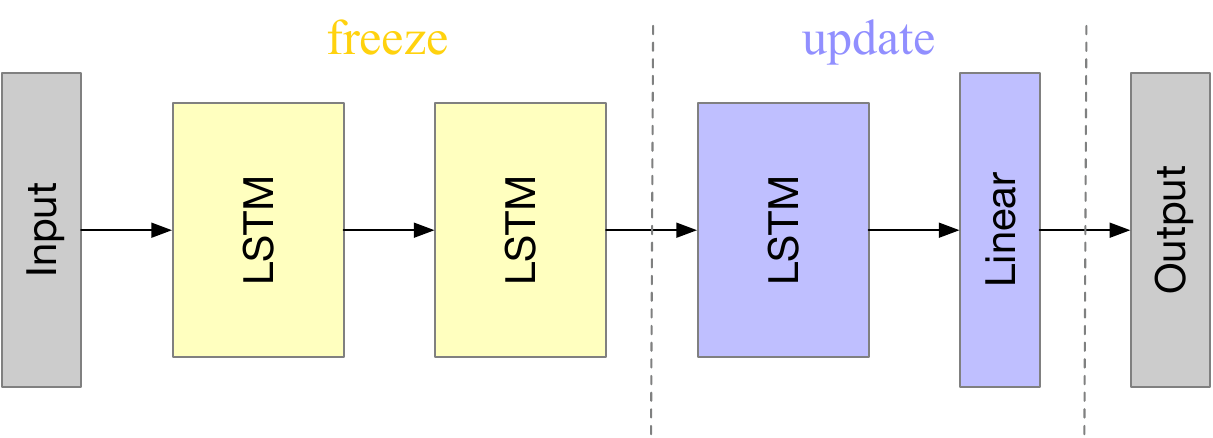}
    \caption{}
    \label{fig:tl_fe_arch}
    \end{subfigure}
    \hfill
    \begin{subfigure}[b]{0.5\columnwidth}
    \includegraphics[width=\columnwidth]{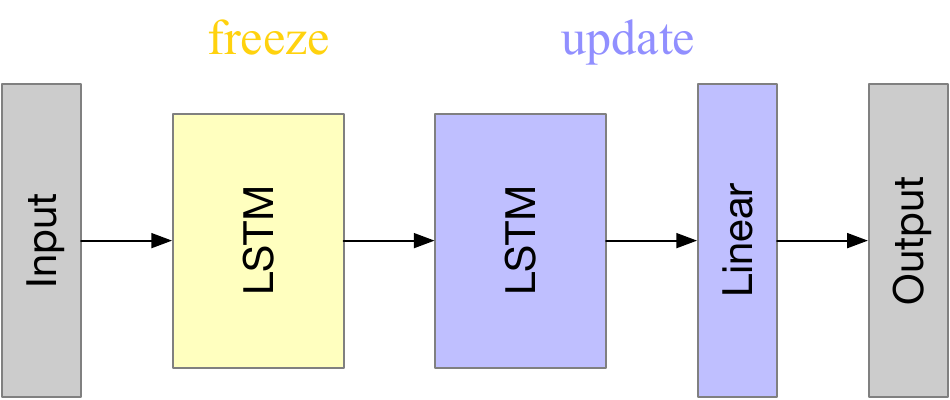}
    \caption{}
    \label{fig:tl_ft_arch}
    \end{subfigure}
    \caption{Next location prediction architectures: (a) traditional general model; (b) transfer learning-based feature extraction model; and (c) transfer learning-based fine tuned model.}
    \label{fig:architectures}
    \vspace{-2mm}
\end{figure}

\subsubsection{LSTM-based General Model}
\label{sec:general_model}
The traditional approach has been to use historical trajectories, temporally extended sequences of locations, of many different users to train a deep neural network that predicts the next location of any given user. Early approaches were based on RNNs while the state-of-the-art approaches use LSTMs \cite{hochreiter1997long} to capture both the short-term and long-term dependencies in user mobility patterns. 
%We refer the reader to Hochreiter et al. \cite{hochreiter1997long} for more details on LSTMs. 
Figure \ref{fig:traditional_arch} illustrates an example architecture of a LSTM model with two LSTM layers followed by a linear layer. Since training deep models, including LSTMs, requires a large amount of training data, a common approach is to use historical trajectories of many users to train an accurate but general model that performs next location prediction \cite{song2016deeptransport, kong2018hst, zhang2018deeptravel, zhao2020go, trivedi2020empirical}. %\cite{song2016deeptransport, du2016recurrent, liu2016predicting, lin2017deep, feng2018deepmove, liao2018predicting, jiang2018deep, kong2018hst, zhang2018deeptravel, zhao2020go, feng2020pmf}.

\subsubsection{Personalized Models}
\label{sec:simple_personalization}
While a general LSTM model can learn correlations in mobile behavior across users and perform well across a range of users that behave similarly, they are less effective for individual users who exhibit idiosyncratic or dissimilar behavior. To address this issue, researchers have proposed to train personalized models for users to capture their unique behavior \cite{feng2020pmf}. While a single model is used for all users in case of a general model, personalization requires that a unique model is learned and deployed for each user.

A LSTM model similar to a general model can be used for training personalized models. In this case, historical trajectories {\em from a single user} are used to train each model. The advantage of model personalization is that it can yield more accurate user-specific predictions. However, deep learning models require a large amount of single user data to train each personalized model (unlike a general model where less single user data suffices due to the availability of training data from many similar users). 

\subsubsection{Transfer Learning-Based Personalization}
\label{sec:tl_personalization}
In our work, we assume a different approach for model personalization that overcomes some of the limitations of the above methods. Our approach involves first training a general model for next location prediction using training data from multiple users. Then it uses transfer learning to personalize the general model for a new user using their historical data. The advantage of personalizing an already trained general LSTM model using transfer learning is that it requires less single user historical data than training one from scratch. 
% Personalized models for context-aware applications in the mobility domain are constrained due to three limitations: (1) little user data leading to possibilities of model overfitting and consequently privacy leakage \cite{yeom2018privacy}, (2) skewed distributions of mobility (i.e. individuals travel to some places like home and work more often than a mall), and (3) limited computing resources on mobile phones. We employ transfer learning-based methods to train personalized ML models. Particularly, we first train a general multi-user LSTM on a powerful graphics processing unit (GPU) (typically on the cloud), and then employ the general model along with private user data to learn a personalized model using transfer learning methods. The assumption is that the personalized training will occur on resource-constrained mobile devices to prevent any leaks of sensitive information during training.

% \subsubsection{Data Transformation}
The goal of transfer learning is to transfer knowledge learned from solving one task to assist another \cite{tan2018survey}. Existing areas that employ transfer learning, such as computer vision and natural language processing, typically have a fixed domain size between source and target tasks. However, the domain of the multi-user model can differ from the domain of the single user data for next location prediction. For instance, a general mobility prediction model that is trained for New York City will have a different domain from a user who lives in Boston. 
%However, the emphasis of this work is in evaluating transfer learning methods for location-aware mobile applications, not the development of general models that are valid across domains. Employing heterogenous methods transfer learning methods for mobility is a direction for future work. 
In this work, we assume that the target single-user domain is a subset of the source multi-user domain. Assume the source domain is $D_s$ and target domain is $D_t$, where $D_t \subseteq D_s$. Prior to applying transfer learning, we transform the target data by extending the domain with $D_s - D_t$. In our case, this implies introducing new categories (e.g., $D_s - D_t$) to the existing one-hot encoded location categories in the target data. This simplifies the transfer learning process by equalizing the source and target domains. Employing heterogeneous transfer learning methods for mobility is a direction for future work. 

There are two popular methods for transfer learning, either of which can be used to personalize a general model using a small amount of user data.

% \subsubsection{Transfer Learning-Based Personalization}
% Deep learning models use inductive learning to learn a mapping between input features and output classes. In our work, we assume the general model is a two-layer LSTM with dropout followed by a fully-connected linear layer to get the final output (see Figure \ref{fig:traditional_arch}). The choice of the general time-series model is arbitrary. The general model takes as input sequences of historical trajectories and outputs the next location. Transfer learning for deep learning models aims at utilizing the inductive biases of the source model to assist in solving the target task. The assumption is that the target user's mobility patterns are a subset of the patterns represented by the general population used to train the source model.

% To generate personalized models from small user data, we evaluate two popular transfer learning approaches for deep learning.

\textbf{Feature Extraction.} One popular method to conduct transfer learning is to employ the general model as a feature extractor for learning the generic patterns relevant to the task. The layered architecture of deep learning models assist in learning different features or correlations within the data at different layers. Since the general model takes as input the trajectories of many users, it learns a representation of the generic mobility pattern of the users. The intuition behind feature extraction is to exploit and build on top of the representation learned by the generic model. This is conducted by using the primary representation layers of the trained general model (e.g., first two LSTM layers in Figure \ref{fig:traditional_arch}) and adding a surplus layer or a new shallow model before the final linear layer to learn specific patterns from the single user data. This method requires re-training the model with single-user data, but only updating the parameters of the newly appended shallow model. To ensure that only the newly appended shallow model is updated and the generic patterns are not lost during the training process, the weights of the general model layers prior to the shallow model are frozen. In our work, we stack another LSTM layer before the output layer to capture the patterns unique to the user as shown in Figure \ref{fig:tl_fe_arch}. %Through this, we build on the generic patterns from the general model and capture the unique user patterns using the new LSTM layer. 

\textbf{Fine Tuning.} Another popular transfer learning approach considers fine tuning the trained general model instead of building on top of it. The initial layers in a deep learning model often focus on generic patterns and the latter layers focus on specific patterns relevant to the task at hand. 
%For instance, a next prediction mobility model can learn high-level mobility patterns of users first such as the relationship between locations and time of day, and then focus on particular relationships between the locations such as home and work place.
During transfer learning, the goal typically is to transfer the generic features and learn the specific patterns based on the target data (e.g., single-user trajectory). To do so, one method is to freeze the initial layers and re-train the latter layers with single user data. Figure \ref{fig:tl_ft_arch} shows an example of such a model. The particular number of layers to re-train or fine-tune depends on the nature of the data. With plenty data, more layers can be re-trained whereas with sparse data, often the case with single-user trajectories, minimizing this number can be better due to the risk of overfitting. In our work, we re-train and update parameters of the second LSTM layer and linear layer using single user data. 

%Section 3B
\subsection{Privacy Attacks on Personalized Models}
\label{sec:attack}
As noted in Section \ref{sec:background}, ML models are vulnerable to privacy attacks. A particular type of ML privacy attack is a model inversion attack that exploits a trained ML model to infer values of sensitive attributes in the training data \cite{fredrikson2015model}. While inversion attacks have been studied in other contexts, prior work has not explored inversion attacks on time-series based ML models, and specifically, context-aware services that use time-series trajectories of contexts such as location history.

Intuitively, a model inversion attack takes a trained model as a black box and a possible output $y$ produced by the model to learn one or more model input features that produce this output. A simple model inversion attack exploits confidence values and prior knowledge of the sensitive variable by picking the value that maximizes the model's confidence of $y$ weighted by the prior \cite{fredrikson2015model}. In case of a next location prediction model, it implies taking a predicted next location (model output) to learn one of the previous locations (model input) visited by the user. This is concerning given the sensitivity of location data (e.g., visit to a hospital can leak privacy). The goal is to not reveal more than needed for the service to operate.

Model inversion attacks have greater privacy implications for personalized models than general models. Since a general model is trained using data from many users, leaking a previously visited location as present in the training data may not directly reveal private information of a specific user. However, an inversion attack on a personalized model directly reveals prior locations visited by a specific user, which can leak sensitive information about that user. In the rest of this section, we formalize and describe a model attack inversion attack on personalized time-series next-location models.

\subsubsection{Threat Model}
\label{sec:threat_model}
We consider a system which consists of a location-aware mobile application that collects sequences of data $x_t = \{f_1, f_2,...,f_k\}$ with $k$ features at each time step $t$. 
%This application trains a multi-user ML model, $M_G$, for next location prediction on a centralized server (e.g., cloud) using historical sequences and deploys this general model to each new user's local device to build personalized models. Since the personalized training is assumed to be on a local device, the training process is privacy-preserving. 
This system consists of the following entities:

\textbf{Contributors.} We assume there exists a set $\mathcal{G}$ of unique users who allow their data to be used to train a multi-user ML model, $M_G$, for next location prediction. These users serve as \textit{contributors} for $M_G$. % and may or may not have access to $M_G$. 

\textbf{Users.} We consider a set of honest unique users $\mathcal{P}$, disjoint from $\mathcal{G}$, that use the location-aware application. %and have local devices capable of handling AI-intensive tasks (e.g., via edge accelerators). 
We assume all users in $\mathcal{P}$ employ a transfer learning-based personalization method (see Section \ref{sec:tl_personalization}) and general model $M_G$ to build personal models. These users protect their data by keeping it local and only allowing the service provider black-box access to the personal ML model. %personal ML model by only allowing black-box access to the service provider. %These users protect their data by keeping it local and only deploying their trained personal ML model to the centralized server.

\textbf{Service Provider.} We consider a service provider $\mathbb{S}$ that hosts the location-aware mobile application. $\mathbb{S}$ has access to the data sequences of users in set $\mathcal{G}$ using which it trains $M_G$, and only black-box access to trained personal models of users in $\mathcal{P}$. We assume $\mathbb{S}$ has the ability to query and observe the model output and associated confidence scores for all classes. We consider $\mathbb{S}$ to be a honest-but-curious adversary that attempts to learn historical mobility pattern of users in $\mathcal{P}$ using their personal ML models. 

% \textbf{Outsiders.} We also consider adversaries that are external to the above mentioned entities. We assume the system established by $\mathbb{S}$ is secure. How $\mathbb{S}$ protects the system from external adversaries is out of the scope of this paper.
%Outsiders could have access to the personal ML models of users in $\mathcal{P}$. We assume such adversaries can observe a limited set of locations of the user via other means, such as other context-aware applications, mobile cookies, third-party applications or location-based social networks. Through the model and a small set of prior user locations, outsiders can attempt to derive the historical mobility pattern of users in $\mathcal{P}$, similar to $\mathbb{C}$.

Since our focus is on privacy rather than security, we do not consider security threats from external adversaries who may break into the system and steal private data or models.

\subsubsection{Proposed Privacy Attack}
Our focus in this paper is on attribute inference attacks using model inversion. The proposed model inversion attack follows the basic premise as described earlier. We assume that all personal models output confidence scores (probabilities) for all classes. This is a typical assumption in mobility applications, particularly when the focus is on getting the top $k$ most likely next locations rather than a single next location. Let $p = (p_1,...,p_m)$ be the marginal probabilities of the sensitive variable that can take $m$ values. For instance, if the sensitive variable is building-level location, the marginal probability $p_i$ will reflect how often building $i$ is visited. 
%In Section \ref{sec:exp_pmethod}, we demonstrate the impact of varying levels of preciseness of $p$ on the attack accuracy. 
The novelty in our work arises from the formalization of this attack from a time-series context. 

\begin{table}[]
\centering
\caption{Descriptions of different adversaries with the components they have access to and their goal. $M_p$ refers to a user's personalized model, $p$ refers to prior knowledge, $x_{t-1}$ and $x_{t-2}$ are inputs to $M_p$, and $l_t$ is the output of $M_p$.}
\begin{tabular}{@{}lcccccc@{}}
\toprule
\textbf{Adversary}  & \multicolumn{5}{c}{\textbf{Adversarial Knowledge}}             & \textbf{Goal} \\
                    & $M_P$      & $p$        & $x_{t-1}$  & $x_{t-2}$  & $l_t$      &                \\ \midrule
A1                  & \checkmark & \checkmark & -          & \checkmark & \checkmark & $l_{t-1}$ \\
A2                  & \checkmark & \checkmark & \checkmark & -          & \checkmark & $l_{t-2}$ \\
% A3                  & \checkmark & \checkmark & -          & \checkmark & -          & $l_{t-2}$ \\
% A4                  & \checkmark & \checkmark & \checkmark & -          & -          & $l_{t-1}$ \\
A3                  & \checkmark & \checkmark & -          & -          & \checkmark & $l_{t-1}$ or $l_{t-2}$ \\
%A4                  & \checkmark & -          & -          & -          & \checkmark & $l_{t-1}$ or $l_{t-2}$ \\ 
%A5                  & -          & \checkmark & -          & -          & \checkmark & $l_{t-1}$ or $l_{t-2}$ \\ 
\bottomrule
\end{tabular}
\vspace{-2mm}
\label{tab:adversaries}
\end{table}

We assume that adversarial access to features is limited by time. That is, an adversary has access to all or no features within a sequence for a given time step. For simplicity, we further assume that there is a single sensitive variable at each time step (e.g., location $l$) for all adversaries. Table \ref{tab:adversaries} presents descriptions of different adversaries with the features they have access to and their goal. We assume all adversaries have access to some location of the user. A honest-but-curious service provider can simply observe the output of the personal models (i.e., $l_t$) or gather such information from other context-aware applications, mobile cookies, third-party applications or location-based social networks. A1 and A2 represent the simplest adversaries which have access to all features except features at time $x_{t-1}$ or $x_{t-2}$ with the goal of correctly identifying $l_{t-1}$ and $l_{t-2}$ respectively. Note, these adversaries require some historical external information namely all features at time $t$--$2$ and $t$--$1$ respectively. 
Adversary A3 represents an adversary who has limited access to historical sequences but has information on model output or some location $l_t$.% Lastly, A4 represents a baseline adversary that does not require marginal probabilities $p$ to access historical mobility patterns.

A popular form of model inversion attacks require enumeration over values of the sensitive variable(s). %In the \textit{optimal} scenario and often with non time-series data, only non-sensitive variables are unknown and the adversary has access to all other features. 
The simplest and most computationally expensive form of enumeration for time-series data is a \textit{brute force} method where an adversary enumerates through all the features in an unknown sequence $x_t$. Since deep learning models learn a differentiable mapping between the input and the output, it is also possible to reconstruct the input using the output through backpropagation and \textit{gradient descent}. Backpropagation is used in deep learning to calculate the gradient of the loss function with respect to the parameters of the model and gradient descent allows a descent or step in the direction that optimizes the loss function through the gradient. We employ this algorithm to reconstruct the input, sequences $x_{t-2}$ and $x_{t-1}$, by iteratively transforming a candidate input towards the values that maximize the correct output. To deal with the large output space typical in mobility domains, we also add the notion of temperature scaling. Temperature, $\mathcal{T}$, is a hyperparameter that controls the variability in prediction space by scaling the raw probabilities (i.e., logits) before applying softmax. The logits ($z_i$) are divided by this term before applying the softmax function:
\begin{align} 
    p_i =  \frac{\exp{(z_i/\mathcal{T})}}{\sum_i \exp{(z_i/\mathcal{T})}} 
    \label{eq:soft_temp}
\end{align}
We use this as a method to soften the candidate input variables during gradient descent such that they are one-hot encoded and represent discretized values.

% \begin{figure}
%     \centering
%     \includegraphics[width=0.8\columnwidth]{figures/inversion_attack.png}
%     \caption{Placeholder for inversion attack.}
%     \label{fig:inversion_attack}
% \end{figure}

Additionally, we propose an enumeration method that employs the \textit{time-based dependence} between the features. Considering that mobile devices are consistently with users, we can assume that there exists cross-correlation between consequent sequences and continuity (e.g., no gaps in time periods).  Thus, we can use smart enumeration techniques that take advantage of these correlations by enumerating through only certain features and using cross-correlation to infer the rest. This method is dependent on the nature of the input features and works for numerical time-varying features. For example, if we assume a sequence consists of location ($l$), duration at location ($d$), and entry time at location ($e$), for adversary A1, we can enumerate through $d_{t-2}$ and $l_{t-2}$ and compute $e_{t-2}$ from knowledge of $e_{t-1}$ and $d_{t-2}$ (e.g., $e_{t-2} = e_{t-1} - d_{t-2}$). Moreover, to minimize the search space, we propose identifying the user's locations of interest. Since the adversary is assumed to have black-box access to the model, we propose observing the output for a few instances and selecting only locations with confidence greater than or equal to some threshold (i.e. 1\%). This will minimize the search space substantially, particularly since the personalized model includes all locations in a given proximity, instead of only those captured in the user's data due to the domain equalization mentioned in Section \ref{sec:tl_personalization}.

%Mention GD-based method?
%Section 5
\section{Privacy Leakage from Inversion Attacks}
In this section, we empirically evaluate the efficacy of the model inversion privacy attack presented in Section \ref{sec:attack}.

\subsection{Experimental Setup}
\label{sec:exp_setup}

\textbf{Data.} We employ a campus-scale WiFi dataset from September to November 2019. This data consists of 156 buildings that are connected by 5104 HP Aruba access points (APs). Each AP event includes a timestamp, event type, MAC address of the device and the AP. Since the WiFi network requires all users to authenticate themselves, each event can be associated with a user. For this work, all user information is anonymized using a hashing algorithm.

Using well known methods for extracting device trajectories from WiFi logs (e.g., \cite{trivedi2020empirical}), we extract fine-grained mobility trajectory of 300 users spanning over 150 buildings and 2956 APs. We filter the data to consist of only on-campus students by assessing whether users stay in a dorm on a typical weekday night. The final processed data set includes sequences of four features for each user: \textit{session-entry} ($e$), \textit{session-duration} ($d$), \textit{building} ($l$), and \textit{day-of-week} ($w$). Note, \textit{session-entry} is discretized into 30 minutes intervals and \textit{session-duration} is discretized into 10 minutes intervals to reduce the variability. Duration is also capped at 4 hours since less than 10\% of users spend more time in a single building \cite{trivedi2020empirical}. %Table \ref{data_description} provides detailed statistics of 200 randomly selected users post processing.

% \begin{table}
% \centering
% \begin{tabular}{ l c } 
% \hline
% \textbf{Item Description} & \textbf{Statistics}  \\
% \hline
% Number of Users & 200 \\ 
% Number of Buildings & 150 \\ 
% Number of APs & 2956 \\ 
% %Number of Devices & 477 \\ 
% %Avg. Devices Per User & 2.40 \\ 
% Avg. Building Visits Per User & 3.26/day \\ 
% Avg. AP Visits Per User & 5.10/day \\ 
% Avg. Time Spent in Building & 46.19 mins \\
% Avg. Time Spent in AP & 32.50 mins \\
% \hline
% \end{tabular}
% \caption{Statistics of the campus-scale trajectories.}
% \label{data_description}
% \end{table}

\textbf{Task.}
We focus on next-location prediction using historical trajectories. Let $x_t = [e_t, d_t, l_t, w_t]$ be a sequence at time $t$. Then, let the ML model be $M: x_{t-2}, x_{t-1} \to l_t$. That is, the ML model takes as input two sequences and outputs the next location. We employ both building-level and AP-level spatial scales for our experiments. 
%We also include a third multi-output model which predicts both building-level and AP-level locations. 
Location $l$ is considered to be a sensitive variable.

\textbf{Models.} We employ trajectories of 200 users to train the general LSTM as described in Section \ref{sec:general_model}. 80\% of the data is used for training and 20\% is used for testing. We perform grid search on time-series based 5-fold cross validation to select the optimal hyperparameters for the model. The general LSTM is trained using a learning rate of $1\mathrm{e}{-4}$ with a weight decay of $1\mathrm{e}{-6}$ and hidden layer size of $128$. We use batches of size $128$ with a dropout rate of $0.1$ between the LSTM layers. 
%To learn personalized models, we employ four methods on personal user data: 
% \begin{enumerate}
%     \item \textbf{Reuse:} reusing the general model without modifications to do personalized predictions (baseline)
%     \item \textbf{LSTM:} training a 1-layer LSTM with dropout
%     \item \textbf{TL Feature Extract (FE):} employing transfer learning-based feature extraction on the general model %(see Section \ref{sec:personalization} for details)
%     \item \textbf{TL Fine Tune (FT):} employing transfer learning-based fine tuning on the general model %(see Section \ref{sec:personalization} for details)
% \end{enumerate}
To learn personalized models, without loss of generality, we employ transfer learning-based feature extraction (TL FE) (see Section \ref{sec:tl_personalization}). 
%Without loss of generality, all attacks are executed on the TL FE personalized model. 
We train individual personalized models for 100 unique and distinct users. We perform grid search using 3-fold time-series cross validation for hyperparameter selection.

\textbf{Measures.} We employ \textit{top-$k$ accuracy} as an evaluation metric. The goal is to identify the top-$k$ most likely locations from the model output and assess whether the true location is a subset of that.

\subsection{Analysis of Privacy Attack}
\label{sec:analysis_attack}

\begin{figure*}
    \centering
    \begin{subfigure}[b]{0.29\linewidth}
    \includegraphics[width=\linewidth]{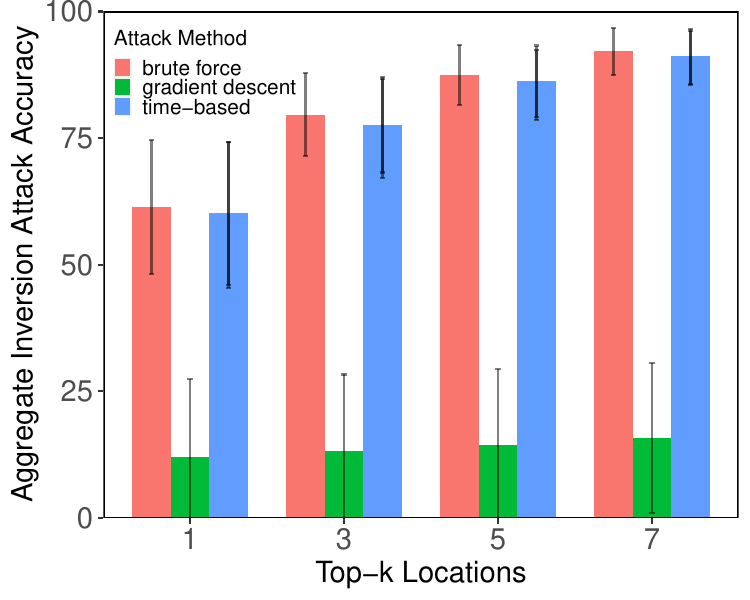}
    \caption{}
    \label{fig:exp_enum_method}
    \end{subfigure}
    % \hfill
    %
    \begin{subfigure}[b]{0.29\linewidth}
    \includegraphics[width=\linewidth]{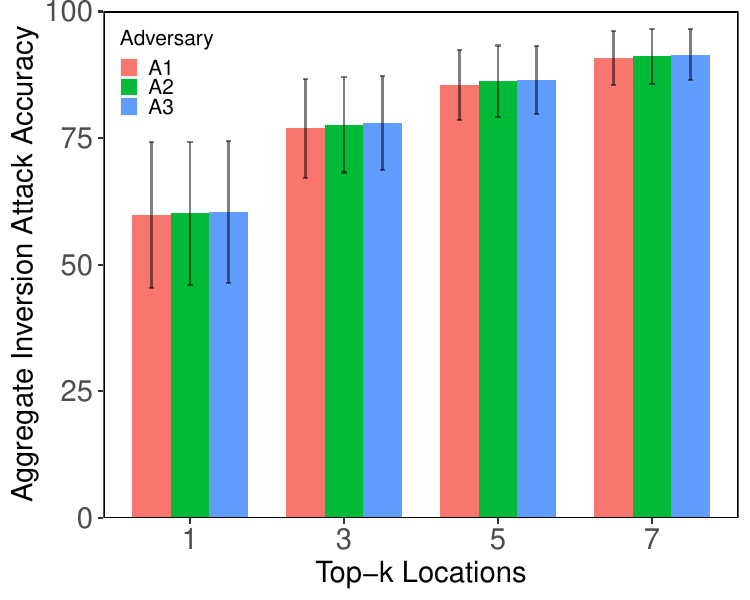}
    \caption{}
    \label{fig:exp_adv_knowledge}
    \end{subfigure}
    % \hfill
    %
    \begin{subfigure}[b]{0.29\linewidth}
    \includegraphics[width=\linewidth]{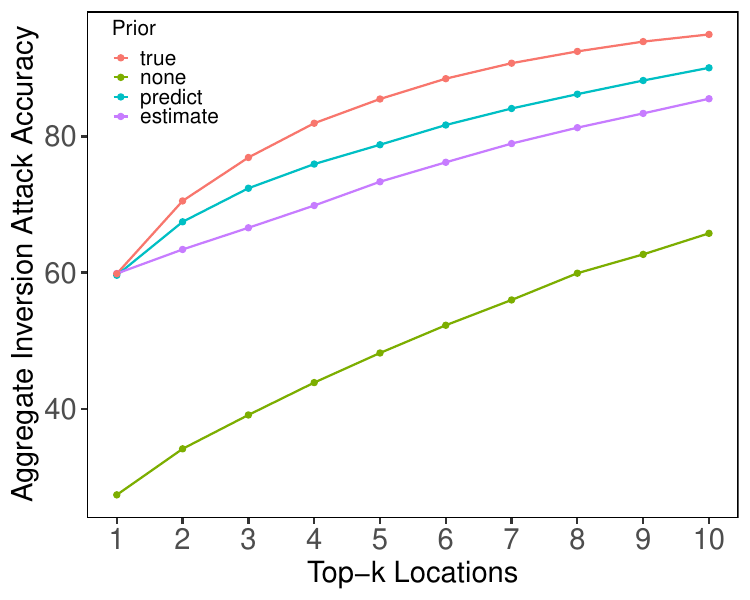}
    \caption{}
    \label{fig:exp_marg_prob}
    \end{subfigure}
    \caption{Results of evaluating the efficacy of the privacy attack under varying system configurations: (a) impact of varying attack methods; (b) impact of varying adversarial knowledge; and (c) impact of nature of prior knowledge $p$.}
    \label{fig:attack_exp1}
\end{figure*}

We analyze the proposed privacy attack on 100 distinct users. We use time-based dependence and adversary A1 as our default attack method and adversary respectively, and perform all experiments on building spatial level unless otherwise stated. For all experiments, attack accuracy is defined as the percentage of historical locations correctly identified.

% \begin{figure}
%     \centering
%     \begin{subfigure}[b]{0.49\columnwidth}
%     \includegraphics[width=\columnwidth]{figures/s2_tl-feat_exp_enum_bar_bldg.png}
%     \caption{}
%     \label{fig:exp_enum_method}
%     \end{subfigure}
%     \hfill
%     %
%     \begin{subfigure}[b]{0.49\columnwidth}
%     \includegraphics[width=\columnwidth]{figures/s2_exp_pm_bldg.png}
%     \caption{}
%     \label{fig:exp_pm}
%     \end{subfigure}
%     \caption{Impact of varying enumeration methods and personalization methods on privacy respectively.}
%     \label{fig:attack_exp1}
% \end{figure}

\subsubsection{Impact of attack type} %enum method: brute, optimal, time, gd, none
We compare the two proposed attack methods, time-based enumeration and gradient descent with temperature scaling, with brute force. Figure \ref{fig:exp_enum_method} contains an evaluation of the attack methods discussed in Section \ref{sec:attack}. As expected, the brute force method performs well, reaching 79.64\% attack accuracy for top-3 predictions. Our proposed time-based method performs equivalently to the brute force method with attack accuracy growing as $k$ increases. However, the gradient descent method is the least effective at constructing historical mobility patterns with attack accuracy of less than 16\%. We hypothesize this is due to the large domain size and discrete nature, instead of continuous, of mobility locations which results in an inaccurate reconstruction of the historical data.

Despite the similar performance, the brute force and time-based enumeration methods differ substantially in computational complexity. The runtime of the brute force method is over 120 times that of the time-based method suggesting that the time-based attack is highly efficient to launch in practical settings. Table \ref{tab:runtime} contains runtimes of the three methods. 

% \textit{\textbf{Takeaway:} The proposed time-based enumeration method gives the best privacy attack accuracy and computational efficiency when compared to the brute force and gradient descent methods. The gradient descent method is the least effective at reconstructing historical mobility patterns.}
%time: 15003 seconds --> 4.17 hours

% \subsubsection{Impact of personalization method} 
% We also consider the impact of personalization methods described in Section \ref{sec:exp_setup} on attack effectiveness. The results in Figure \ref{fig:exp_pm} indicate that all personalization methods have similar privacy leakage. Despite these methods having differing model accuracies, their privacy leakage is similar with the proposed transfer learning methods having slightly higher attack accuracy.

% \textit{\textbf{Takeaway:} All personalization methods (reuse, LSTM, TL FE, and TL FT) have a similar privacy leakage.}

\begin{table}
\centering
\caption{Runtime of attack methods for 100 users.}
\begin{tabular}{ l c } 
\hline
Method & Runtime (hours) \\
\hline
Brute Force & 82.18 \\ 
Gradient Descent & 6.27 \\ 
Time-Based & 0.68 \\ 
\hline
\end{tabular}
\vspace{-2mm}
\label{tab:runtime}
\end{table}

\subsubsection{Impact of adversarial knowledge} %A1 to A4
The results shown in Figure \ref{fig:exp_adv_knowledge} illustrate the impact of adversarial knowledge from Table \ref{tab:adversaries} on the attack. Despite the differing levels of adversarial knowledge, all adversaries perform effectively and equivalently at reconstructing historical mobility patterns. Interestingly, adversary A3's attack capabilities do not degrade despite the lack of adversarial knowledge. This illustrates that even with limited prior information on historical time steps, an adversary can effectively perform a model inversion attack.

% \textit{\textbf{Takeaway:} Adversarial knowledge of historical time steps does not affect the attack efficiency; all adversaries presented in Table \ref{tab:adversaries} form equivalently.} 

\subsubsection{Impact of prior information} %p method 
\label{sec:exp_pmethod}
All experiments thus far assume that the adversary has access to the true marginal probabilities of the sensitive variable. However, this is unlikely to be known by a typical adversary. In reality, an adversary can get access to the most probable value(s) of the sensitive variable but not know exact probabilities. We attempt to \textit{estimate} the marginal probabilities $p$ in this manner by assigning a high probability (e.g., 75\%) to the most probable value and equally distributing the remaining probability among the other values. The adversary can also easily observe the output of the target model for a period of time and \textit{predict} $p$. Figure \ref{fig:exp_marg_prob} demonstrates the impact of different $p$ generation methods, namely \textit{true}, \textit{none}, \textit{predict} and \textit{estimate}. 

The results in Figure \ref{fig:exp_marg_prob} confirm the importance of using $p$ during the attack; without $p$, the attack is less effective. However, the attack is not sensitive to the precision of $p$. The \textit{true} method results in the highest attack effectiveness across $k$ whereas predicting or estimating $p$ results in a 5-10\% degradation in attack efficacy. The difference between true, predict and estimate methods grows as $k$ increases. Naturally, among these three, the effectiveness of the estimate method grows the slowest as $k$ increases, due to its highly skewed probability estimates. 

% \textit{\textbf{Takeaway:} Attacks perform poorly without accurate estimates of the marginal probability $p$. However, attacks are not sensitive to the precision of $p$. Estimating or predicting $p$ results in 5-10\% loss in effectiveness suggesting that predicting $p$ is a reasonable method to conduct the attack when true $p$ is not known.}

\subsubsection{Impact of mobility spatial levels}
Mobility spatial levels (the spatial resolution) can differ based on the task definition. Thus far, all experiments were evaluated at a building-level scale. To understand the impact of a fine-grained spatial scale, we run the attack at the scale of access points (APs). There are 2956 APs in our data set. 

The results in Figure \ref{fig:exp_spatial} show that the attack leaks less privacy at the AP scale when compared to building scale. We hypothesize this is due to the large domain size of AP-level models, which makes it difficult to reconstruct historical patterns. Similar to building scale, there is more privacy leakage as $k$ grows. In future work, we would like to consider ways to handle larger spatial scales.

% \textit{\textbf{Takeaway: } Higher-level spatial scales leak less privacy.}

% \begin{figure}[H]
%     \centering
%     \includegraphics[width=0.7\columnwidth]{figures/s2_tl-feat_exp_spatial.png}
%     \caption{Impact of the privacy attack on different spatial scales. Results are demonstrated for two spatial scales with 150 unique buildings and 2956 unique APs.}
%     \label{fig:exp_spatial}
% \end{figure}

\subsubsection{Impact of degree of mobility}
We also evaluate how characteristics of mobility affect privacy leakage. The degree of mobility varies for different users. Highly mobile users visit many locations and less mobile users tend to visit fewer locations during a given time period. For instance, socially active users may physically move around more than their counterparts. We evaluate how degree of mobility affects attack accuracy in Figure \ref{fig:exp_mobility}. 

The degree of mobility has a weak effect on privacy leakage. Since users tend to spend a majority of their time at a single location \cite{trivedi2020empirical}, it is likely that the attack is less affected by the degree of mobility at less visited locations. These results are supported by a regression analysis; the correlation coefficients are weak, $0.337$ and $0.107$ for building and AP level respectively, with statistically significant p-values ($p \leq 0.05$).

% \textit{\textbf{Takeaway: } Privacy leakage has a weak dependence on the degree of mobility for high and low level spatial scales.}

\subsubsection{Impact of mobility predictability} 
We further evaluate the impact of mobility predictability on attack accuracy. Highly predictable users have highly correlated mobility patterns across time and space. We employ the personalized model accuracy as a proxy for mobility predictability. That is, higher model accuracy implies higher predictability of mobility since the model is expected to capture the correlations in the mobility pattern of the user. 

\begin{figure*}
    \centering
    \begin{subfigure}[b]{0.28\linewidth}
    \includegraphics[width=\linewidth]{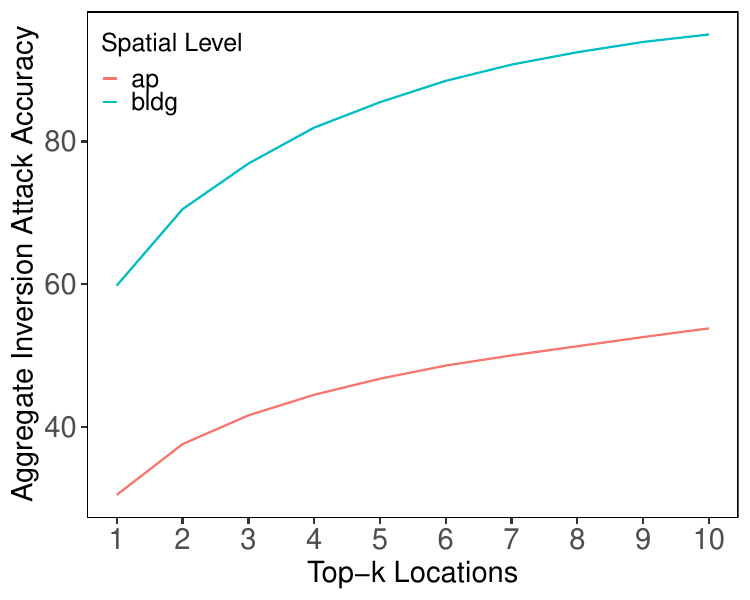}
    % \caption{Impact of the privacy attack on different spatial scales. Results are demonstrated for two spatial scales with 150 unique buildings and 2956 unique APs.}
    \caption{}
    \label{fig:exp_spatial}
    \end{subfigure}
    % \hfill
    %
    \begin{subfigure}[b]{0.28\linewidth}
    \includegraphics[width=\linewidth]{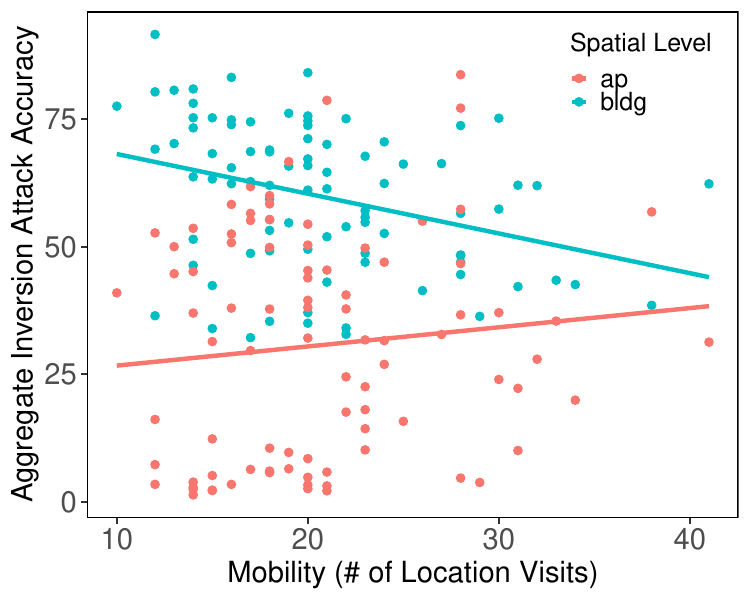}
    \caption{}
    \label{fig:exp_mobility}
    \end{subfigure}
    % \hfill
    %
    \begin{subfigure}[b]{0.28\linewidth}
    \includegraphics[width=\linewidth]{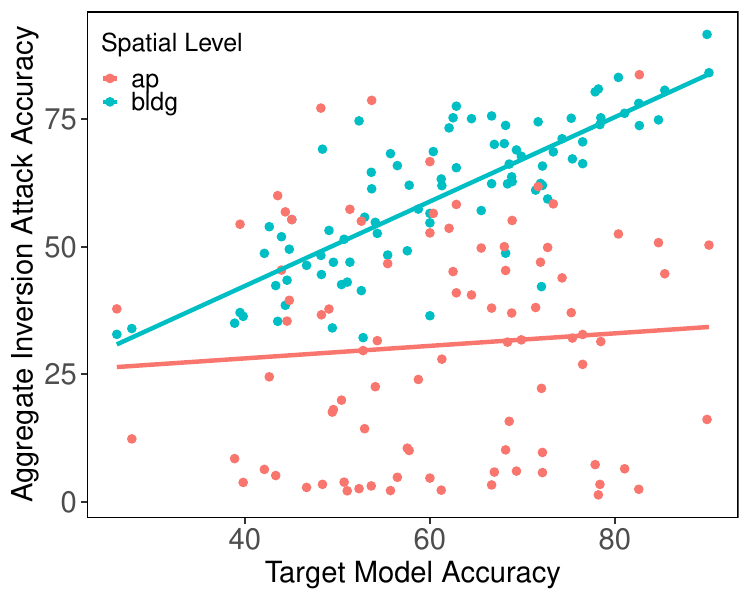}
    \caption{}
    \label{fig:exp_predictability}
    \end{subfigure}
    \caption{Results for two spatial scales with 150 unique buildings and 2956 unique APs: (a) impact of the privacy attack on different spatial scales; (b) impact of degree of mobility on privacy; and (c) impact of mobility predictability on privacy.}
    \label{fig:attack_exp2}
\end{figure*}

We show results in Figure \ref{fig:exp_predictability}. Mobility predictability strongly affects privacy leakage for building spatial level. This is not surprising since the attack is based on inverting the model itself; more accurate models more precisely capture mobility patterns which can then be exploited by the attack. These results are supported by numerical results from regression analysis. There is a strong correlation coefficient of $0.804$ with a statistically significant p-value ($p=2.92\mathrm{e}{-2}$). However, we note that the relationship is weak for AP spatial level with a correlation coefficient of $0.078$ and insignificant p-value of $0.031$. We hypothesize that the distribution of time spent in different APs can explain the variance in attack accuracies for similar target model accuracies seen in Figure \ref{fig:exp_predictability}.

% \textit{\textbf{Takeaway:} Mobility predictability has a strong correlation with privacy leakage for higher-level spatial scales; higher correlations in mobility patterns results in higher privacy leakage. Lower-level spatial scales have a weak correlation with mobility predictability and are likely confounded by the range of motion across APs.}

\textbf{Key Takeaways:} The proposed time-based model inversion attack is computationally efficient and effective at revealing historical mobility patterns with 77.61\% accuracy for top-3 estimates even with limited adversarial knowledge and low precision prior $p$. While the privacy leakage is independent of the mobility behavior of the user, there is a trade-off between model efficacy (i.e. correlation in data) and privacy. Furthermore, models of coarse-grained spatial scales leak more privacy. These results demonstrate that context-aware personalized models can be easily exploited with limited information for users with highly correlated mobility patterns.

%Section 6
\section{Privacy Preserving ML Framework for Mobile Services}
In this section, we present \pelican, a privacy preserving framework for machine learning-based mobile services.

\subsection{System Design}
\begin{figure*}
    \centering
    \includegraphics[width=0.9\textwidth]{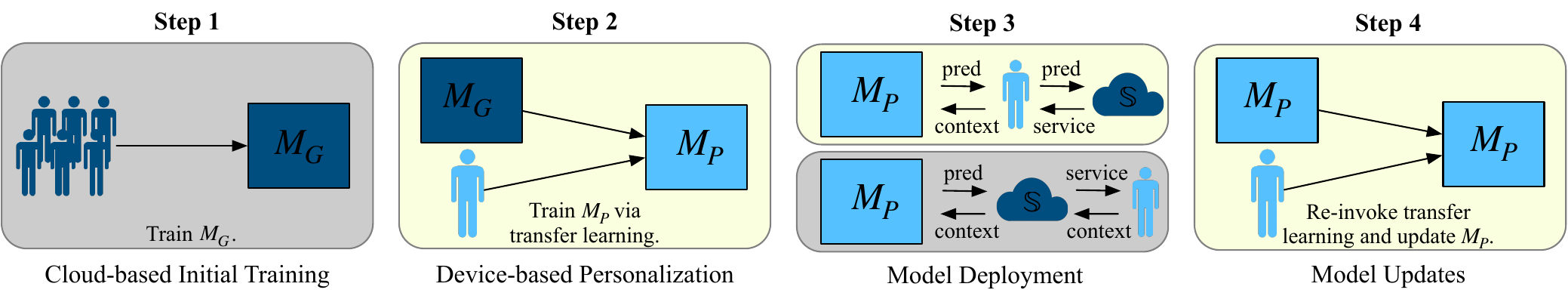}
    \caption{Overview of the proposed system, \pelican. Grey and yellow represents phases that are executed on the cloud and device respectively. Phase 3 can occur in either the cloud or the device depending on the service characteristics. $M_G$ is the general multi-user model, $M_P$ is the personalized model and $\mathbb{S}$ is the service provider.}
    \label{fig:pelican}
\end{figure*}

\pelican is a distributed framework for training and deploying personalized ML models for mobile services in a privacy preserving manner. \pelican's architecture is designed to safeguard private training data of individual users, such as historical location trajectories, while learning a personalized model. \pelican also incorporates privacy preserving enhancements into the deep learning model itself to thwart model inversion attacks. The framework leverages the device and cloud tiers of distributed mobile services to achieve its goals. Figure \ref{fig:pelican} depicts the design of \pelican.

\pelican comprises of the following key components:

\subsubsection{Cloud-based Initial Training}
The first step in designing a privacy-preserving ML model for mobility is to train a general model, $M_G$, using training data from multiple users. Since initial training of the model is compute intensive, this component of our framework runs on cloud servers and leverages specialized resources such as GPUs when available. The initial training components invokes a deep learning library on a cluster of cloud servers to train a general model. For example, in case of next location prediction, we train a LSTM-based deep learning model using time-series trajectories of locations visited by 200 users over a duration of two months.

\subsubsection{Device-based Personalization}
Once a general ML model has been trained in the cloud, the next phase personalizes this model for each user using transfer learning. The personalization involves using a small amount of training data for each new user to learn a distinct personalized model, $M_P$. Since the personal training data contains sensitive private information (e.g., location visits), the training for personalization is executed on the local device rather than the cloud. Retaining all private data on local user-owned devices enhances privacy.

To do so, the general model is downloaded from the cloud to the device and transfer learning is performed on the device using personal training data (e.g., location history of the user).  Transfer learning-based personalization can be conducted via feature extraction or fine tuning (see Section \ref{sec:tl_personalization}) depending on the nature of the data. If the personal training data is sparse, feature extraction should be used to avoid overfitting. Note that unlike training the general model which is compute intensive and is performed in the cloud, transfer learning is much less compute intensive and can be performed on devices that are resource constrained \cite{yoon2017efficient}. This phase also involves adding privacy preserving enhancements to the LSTM model (as discussed in Section \ref{sec:solution}).

\subsubsection{Model Deployment}
Once the model has been personalized using transfer learning, it is ready for deployment in the mobile service. Since mobile services can vary in their characteristics, the model can be deployed in two ways. 

The first approach is local on-device deployment where the model executes on the device for making predictions. This approach is suitable for mobile services that run largely on devices (e.g., smartphone mobile apps with a lightweight cloud component). Local deployment avoids network latency to the cloud for AI inference and ensures that the model stays on the user's device minimizing the amount of information known by the service provider and consequently enhancing privacy.

The second approach is to deploy the personalized model in the cloud. This approach is suitable for cloud-based services and enables the service to invoke the model in the cloud to provide context-aware service to the user. In this case, even though the model runs in the cloud, its privacy enhancements prevent model inversion attacks (see Section \ref{sec:solution}).

\subsubsection{Model Updates}
It is common for production services to periodically re-train the ML model to update it with new training data as it becomes available. In our case, as new personal data becomes available, the transfer learning process can be re-invoked to update the parameters of the personalized model, after which it is redeployed for use by the service.

The framework also allows the general model to be updated in the cloud periodically, but this requires re-running the transfer learning process on the device to re-personalize the model for each user. Due to the higher overheads of doing so, updates to the general model are done infrequently while updates to the personalized model can be done frequently.

\subsection{Privacy Enhancements to Personalized Models}
\label{sec:solution}
We now present our privacy enhancement to the LSTM model during model personalization that is designed to thwart inversion attacks. Our goal is to protect training data privacy such that adversaries cannot reverse-engineer a black-box personalized model to learn historical mobility patterns.

The proposed attack thrives on the adversary’s ability to access the model’s output and confidence scores. The enhancement aims to satisfy the following requirements:

\begin{enumerate}
    \item The personalized model can be accessed by the service provider in a black-box manner. This allows the service provider to query the model.
    \item The service provider can access model outputs to get context-aware predictions. The service provider can also access confidence scores to compute the top-k locations.
    \item The service provider cannot determine historical mobility patterns by reverse engineering the model.
\end{enumerate}

The proposed enhancement is based on modification of the confidence scores such that the attack space reduces tremendously. Our approach introduces a new layer into the LSTM model between the linear layer and softmax layer that changes the distribution of the confidence scores without compromising model accuracy. This layer takes as input the raw probabilities from the linear layer. Before applying the softmax function to normalize these raw probabilities, this layer scales the probabilities by dividing them with a value $\mathcal{T}$. Note, this is similar to using temperature scaling, a single parameter extension of Platt scaling \cite{platt1999probabilistic}, in deep learning. Temperature is a hyperparameter often used to control the randomness in the predictions (see Equation \ref{eq:soft_temp}).  

In our work, we use the notion of temperature as a privacy tuner to change the sensitivity to the different outputs at inference time only. As the temperature tends to 0, the confidence of the sample with the highest probability tends to 1. Intuitively, this makes the attack more difficult because the confidence scores will be highly insensitive (i.e., close to 0 or 1). With sharper confidence values, the attack space will reduce and adversaries will not be able to reconstruct historical mobility patterns meaningfully. Note, since the order of the confidence values do not change during scaling, the model's accuracy will remain unaffected as long as appropriate precision is used in storing the confidence values.

The enhancement is designed as a user-centric mechanism; we use this parameter as a value that can be determined by the user. The user can pick a small or large value depending on how much privacy (i.e., insensitivity to the confidence scores) they prefer. 
%Note, as this value gets smaller and the confidence values tend to 0 and 1, the confidence of the top-k outputs will need to be stored with appropriate precision to not hinder with the accuracy of the model. 
%it will hinder the ability to predict the top-k outputs beyond the most likely output (with a low privacy parameter, the most likely output would have 100\% probability). 
We assume the value of the privacy tuner is kept private from the service provider.

\subsection{System Evaluation}

\subsubsection{Prototype and Experimental Setup}
To evaluate \pelican, we employ the same campus-scale WiFi dataset, next-location prediction task and top-$k$ measure as described in Section \ref{sec:exp_setup}. The system prototype is implemented in Python and all deep learning models are built using PyTorch. 
%We assume the cloud-based initial training is run on a GPU cluster and the device-based personalization is run on an individual machine (e.g., mobile device).
The general model in the cloud-based initial training follows the architecture presented in Figure \ref{fig:traditional_arch} and is implemented using PyTorch's \texttt{nn.LSTM}, \texttt{nn.Dropout} and \texttt{nn.Linear} layers.
The transfer learning-based feature extraction method is implemented using PyTorch's \texttt{nn.Sequential} container, whereas the transfer learning-based fine tune method involved re-training part of the general model. We assume that the model is deployed in a location-aware application. For model updates, we initialize the model parameters to that of the trained personalized model and re-invoke the transfer learning process with more data to update the model's parameters. The privacy enhancement is only employed during inference and does not interfere with the training of the model.

The cloud-based initial training of the general model is performed on a server with a NVIDIA Titan-X GPU with 64GB memory with the same setup as mentioned in Section \ref{sec:exp_setup}. 

For device-based personalization, we compare four methods on personal user data: 
\begin{enumerate}
    \item \textbf{Reuse:} reusing the general model without modifications to do personalized predictions (baseline)
    \item \textbf{LSTM:} training a 1-layer LSTM with dropout\footnote{Since there is limited personal user training data, a single layer LSTM model with dropout is a sufficient baseline.}
    \item \textbf{TL Feature Extract (FE):} employing transfer learning-based feature extraction on the general model %(see Section \ref{sec:personalization} for details)
    \item \textbf{TL Fine Tune (FT):} employing transfer learning-based fine tuning on the general model %(see Section \ref{sec:personalization} for details)
\end{enumerate}
As before, we train individual personalized models for 100 unique and distinct users on a low-end CentOS Linux 7 machine with a 2.20GHz Intel CPU and 8GB RAM. The computing power mimics a resource-constrained mobile device. All personalized models perform grid search using 3-fold time-series cross validation for hyperparameter selection.
% 2.20GHz Intel(R) Xeon(R) CPU E5-2650 v4 and 256 GB RAM. run CentOS Linux 7.
% OS : Linux 7 
% CPU : Xeon E5-2650
% RAM : 256 GB

\subsubsection{Overhead of Model Personalization}
We compare the overheads of the cloud-based initial training and the device-based personalization phases in \pelican with the goal that the latter is much less compute intensive than the former since it runs on mobile devices. Our results demonstrate general model training uses approximately 43,000 billion CPU cycles and takes 4.55 hours, whereas personalized modeling uses on average 15 and 14 billion CPU cycles and takes 6.62 and 5.92 seconds for TL FE and TL FT personalization methods respectively (aggregated for 100 users). These results show that while the general model training requires cloud servers, personalization can be done on low-end mobile or edge devices. %Note as the number of users and depth of the general model are increased, the overhead of training the general model will grow exponentially. 

% general model cycles elapsed: 43,450,415,343,339 (43,000,000,000,000)
% personalized model cycles elapsed (avg of 100 users): 15,858,971,325 (16,000,000,000) (tl fe), 14,182,206,795 (14,000,000,000) (tl ft)

\begin{figure*}[t!]
    \centering
    \begin{subfigure}[b]{0.28\textwidth}
    \includegraphics[width=\textwidth]{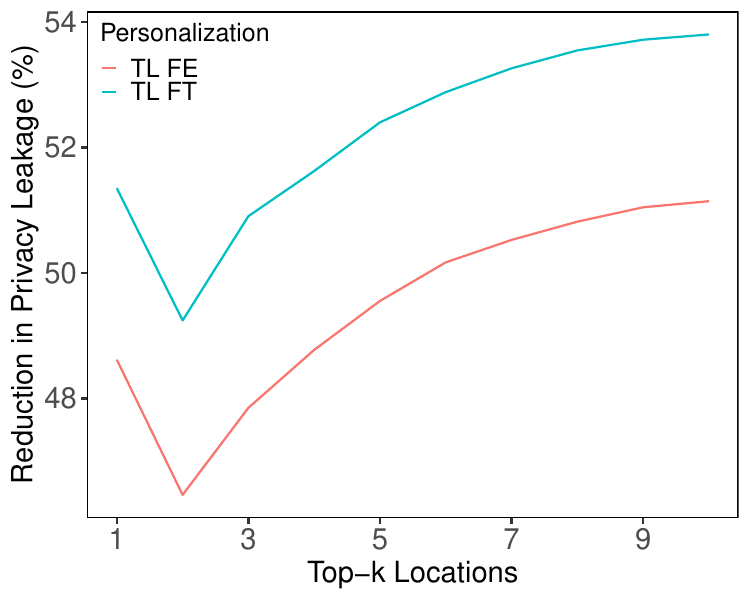}
    \caption{}
    \label{fig:exp_sol_pm}
    \end{subfigure}
    % \hfill
    %
    \begin{subfigure}[b]{0.28\textwidth}
    \includegraphics[width=\textwidth]{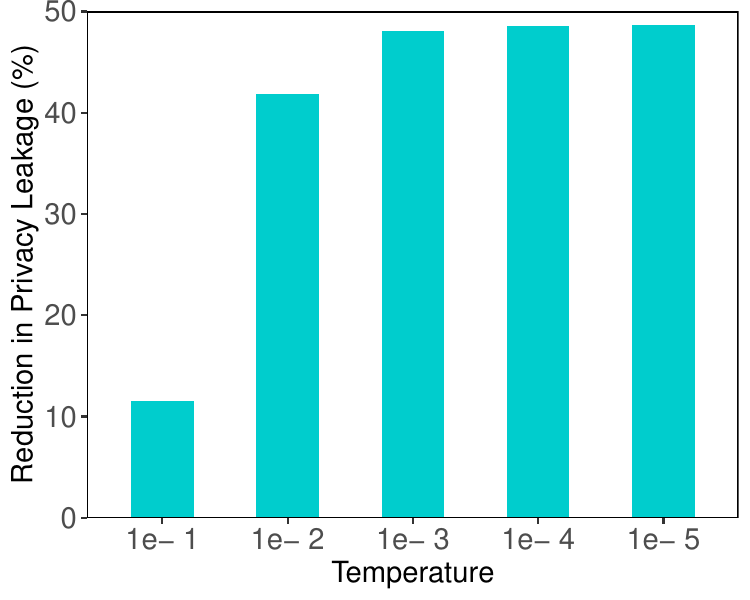}
    \caption{}
    \label{fig:exp_sol_temp}
    \end{subfigure}
    % \hfill 
    %
    \begin{subfigure}[b]{0.28\textwidth}
    \includegraphics[width=\textwidth]{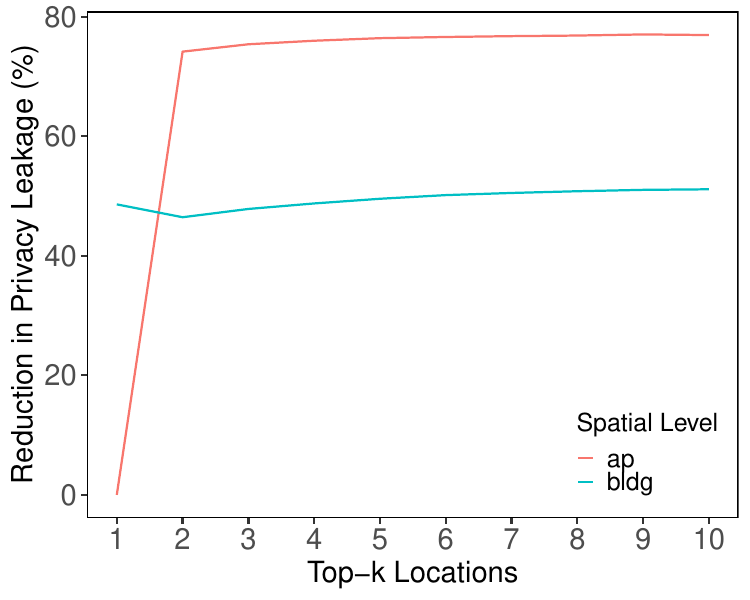}
    \caption{}
    \label{fig:exp_sol_spatial}
    \end{subfigure}
    \caption{Results of the proposed privacy enhancement: (a) impact of the privacy enhancement on personalized models; (b) impact of varying the privacy parameter; and (c) impact of the privacy enhancement on spatial levels.}
    \label{fig:solution_exp}
\end{figure*}

\subsubsection{Efficacy of Device-based Personalization}
\begin{table}
\centering
\caption{Aggregate train and test accuracy (\%) of different personalization methods on 100 individual users. Results demonstrate that transfer learning-based personalization methods employed in \pelican increase test accuracy where the feature extraction (FE) method is least prone to overfitting.}
\begin{tabular}{ c | c | c | c c c } 
\hline
\textbf{Location} & \textbf{Method} & \textbf{Train (\%)} & \multicolumn{3}{c}{\textbf{Test (\%)}} \\
& & & top-1 & top-2 & top-3 \\
\hline
\multirow{4}{*}{Building} & Reuse & 52.16 & 53.02 & 60.09 & 63.68 \\  
 & LSTM & 70.26 & 60.00 & 72.03 & 78.62 \\ 
 & TL FE & 67.81 & \textbf{61.19} & \textbf{72.62} & \textbf{79.05} \\
 & TL FT & 76.47 & \textbf{60.70} & \textbf{73.16} & \textbf{79.61} \\ 
\hline
\multirow{4}{*}{AP} & Reuse & 27.02 & 28.01 & 32.18 & 34.42 \\  
 & LSTM & 51.39 & 44.35 & 57.60 & 63.36 \\ 
 & TL FE & 60.56 & \textbf{48.45} & \textbf{61.94} & \textbf{66.52} \\
 & TL FT & 68.38 & \textbf{47.91} & \textbf{62.26} & \textbf{67.36} \\ 
\hline
% \multirow{4}{*}{Building, AP} & Reuse & 14.09 & 14.28 \\  
%  & LSTM & 46.24 & 44.39 \\ 
%  & TL FE & 52.41 & \textbf{50.94} \\
%  & TL FT & 55.45 & \textbf{50.63} \\ 
% \hline
% \multirow{4}{*}{AP2} & Reuse & 27.77 & 29.48 \\  
%  & LSTM & 55.57 & 43.83 \\ 
%  & TL FE & 50.95 & \textbf{50.51} \\
%  & TL FT & 54.75 & \textbf{50.00} \\ 
% \hline
\end{tabular}
\vspace{-2mm}
\label{tab:personalization_results}
\end{table}

Table \ref{tab:personalization_results} contains the aggregate results of the personalization methods at building and AP-level locations for 100 distinct users. The reuse method performs the worst in both cases. 
%while having a greater impact on AP-level accuracy due to the large domain size (2956 classes). 
From the results, we can conclude that the TL FE method performs the best by almost doubling the baseline accuracy for AP predictions and being less prone to overfitting to the personal data compared to the LSTM and TL FT methods. We define overfitting as the discrepancy between train and test accuracy.

The personalized models in Table \ref{tab:personalization_results} are trained with 8 weeks of personal data (note this is equivalent doing device-based personalization followed by iterative model updates in \pelican). We further examine the efficacy of \pelican with differing training data sizes. As mentioned earlier, one of the advantages of the transfer learning-based approaches employed in \pelican is the ability to use small amounts of training data for learning personalized models. The results of training with differing training data sizes are shown in Table \ref{tab:pelican_train_data} for building-level locations. Both the transfer learning personalization approaches perform similarly with only a slight degrade in performance with smaller training data sizes. However, the TL FT and LSTM methods are prone to overfitting with a higher impact on the LSTM performance.

\begin{table}
\centering
\caption{Aggregate train and test accuracy (\%) of 100 individual users with different training data sizes. Results demonstrate that transfer learning-based personalization methods employed in \pelican are efficient even with less training data whereas the LSTM method is highly prone to overfitting.}
\begin{tabular}{ c | c | c | c c c } 
\hline
\textbf{Train Data} & \textbf{Method} & \textbf{Train (\%)} & \multicolumn{3}{c}{\textbf{Test (\%)}} \\
\textbf{Length} & & & top-1 & top-2 & top-3 \\
\hline
\multirow{3}{*}{2 weeks} & LSTM & 86.76 & 46.92 & 59.67 & 67.44 \\ 
 & TL FE & 67.72 & \textbf{49.87} & \textbf{61.45} & \textbf{67.96} \\
 & TL FT & 72.99 & \textbf{51.29} & \textbf{61.71} & \textbf{68.97} \\ 
\hline
\multirow{3}{*}{4 weeks} & LSTM & 91.63 & 52.16 & 65.57 & 72.73 \\ 
 & TL FE & 68.90 & \textbf{56.64} & \textbf{68.16} & \textbf{74.97} \\
 & TL FT & 78.41 & \textbf{56.83} & \textbf{69.85} & \textbf{76.34} \\ 
 \hline
\multirow{3}{*}{6 weeks} & LSTM & 91.79 & 54.12 & 67.34 & 74.13 \\ 
 & TL FE & 69.03 & \textbf{58.34} & \textbf{70.11} & \textbf{76.72} \\
 & TL FT & 77.73 & \textbf{58.90} & \textbf{71.90} & \textbf{78.37} \\ 
 \hline
\multirow{3}{*}{8 weeks}  & LSTM & 70.26 & 60.00 & 72.03 & 78.62 \\ 
 & TL FE & 67.81 & \textbf{61.19} & \textbf{72.62} & \textbf{79.05} \\
 & TL FT & 76.47 & \textbf{60.70} & \textbf{73.16} & \textbf{79.61} \\ 
\hline
\end{tabular}
\vspace{-2mm}
\label{tab:pelican_train_data}
\end{table}

These results also reinforce the complexity of mobility applications \cite{trivedi2020empirical, feng2020pmf}. Predicting mobility is difficult and varies by the range of user mobility and correlation between mobility patterns. %The goal of \pelican is to build personalized models in a privacy preserving manner for location-aware mobile services. %The goal of our work is to develop a privacy preserving framework for building personalized models and improving personalized mobility models is out of the scope of this paper.

\subsubsection{Privacy Leakage}

We perform an evaluation on the reduction in privacy leakage by applying the enhancements presented in Section \ref{sec:solution} during attacks for the same set of users in Section \ref{sec:analysis_attack}. Without loss of generality, all experiments are performed on adversary A1 using the TL FE personalization method and true $p$ unless otherwise stated. All reported reduction in leakages are aggregated over 100 users.

\textit{Impact of privacy layer on personalized models.}
Results in Figure \ref{fig:exp_sol_pm} show the impact of the attack for transfer learning-based personalization methods. The proposed solution is able to reduce privacy leakage by 46-54\% for transfer learning methods. The reduction in privacy leakage is higher for transfer learning-based fine tuning and decreases as $k$ increases in both types of models. Since the confidence of the most probable location tends to 1 with the privacy enhancement, the attack becomes solely dependent on the prior information for $k=1$. Thus, the reduction in privacy leakage is higher for $k=1$ before decreasing slightly for $k=2$ and increasing again.

\textit{Impact of varying the privacy parameter.}
Results in Figure \ref{fig:exp_sol_temp} demonstrate the impact of changing the temperature (privacy parameter) during inference. As the temperature decreases, the privacy leakage decreases eventually flattening out. Note, this will differ for each user and spatial scales.

\textit{Impact of spatial level.}
Figure \ref{fig:exp_sol_spatial} contains the results of applying the proposed defense mechanism on different spatial levels. As can be noted, the reduction in privacy leakage is higher for low-level spatial scales than high-level spatial scales for $k>1$. For the top-1 prediction, the reduction in privacy leakage is bounded at 0. 

\textbf{Key Takeaways:} \pelican is able to thwart privacy attacks in personalized models with up to 75.41\% reduction in leakage while achieving state-of-the-art performance. The privacy enhancement offers a user-centric design to allow users to control the degree of privacy and lowers the ability of the attack to the extent that it is incomprehensible ($< 40\%$ attack efficacy for top-5 predictions) without compromising on model accuracy. 

\section{Related Work}

\begin{table*}[]
\centering
\caption{Prior work relevant to defending against attribute-inference attacks.}
% \scriptsize
\begin{tabular}{@{}c|lccc@{}}
\toprule
\textbf{Phase} & \textbf{Category} & \textbf{\hfill Edge Friendly} & \textbf{Model I/O Accessible} & \textbf{Personalized Protection} \\
% \textbf{} & \textbf{} & \textbf{Friendly} & \textbf{Accessible} & \textbf{Protection} \\
\hline
\multirow{3}{*}{Data Processing} & Artificial data \cite{triastcyn2018generating, zhang2018differentially} & - & \checkmark  & \checkmark \\
 & Data obfuscation \cite{reza2019privacy, zhao2020tradeoff, zhang2018privacy} & \checkmark & -  & \checkmark \\
 & Light-weight encryption \cite{bost2015machine} & \checkmark & -  & \checkmark \\
 \hline
 \multirow{3}{*}{Training} & Distributed training \cite{hamm2015crowd, shokri2015privacy, he2019model} & \checkmark & - & \checkmark \\
 & Secure enclaves \cite{hunt2018chiron, ohrimenko2016oblivious} & - & \checkmark  & \checkmark \\
 & Differential privacy perturbation \cite{abadi2016deep}  & \checkmark & \checkmark & - \\
 %& Noise induction \cite{mireshghallah2020shredder} & - & \checkmark  & \checkmark \\
 \hline
 \multirow{2}{*}{Inference} & Output perturbation \cite{wu2016methodology, jia2019memguard, yang2020defending} & - & \checkmark  & \checkmark \\
 & \textbf{\pelican (this paper)} & \checkmark & \checkmark & \checkmark \\
\bottomrule
\end{tabular}
\vspace{-2mm}
\label{tab:prior_defenses}
\end{table*}

% \begin{table}[]
% \centering
% \caption{Prior work relevant to defending against attribute-inference attacks.}
% \scriptsize
% \begin{tabular}{@{}P{3.7em}|p{13em}P{2.7em}P{4.5em}P{5.5em}@{}}
% \toprule
% \textbf{Phase} & \textbf{Category} & \textbf{\hfill Edge} & \textbf{Model I/O} & \textbf{Personalized} \\
% \textbf{} & \textbf{} & \textbf{Friendly} & \textbf{Accessible} & \textbf{Protection} \\
% \hline
% \multirow{3}{*}{\shortstack[c]{Data\\Processing}} & Artificial data \cite{triastcyn2018generating, zhang2018differentially} & - & \checkmark  & \checkmark \\
%  & Data obfuscation \cite{reza2019privacy, zhao2020tradeoff, zhang2018privacy} & \checkmark & -  & \checkmark \\
%  & Light-weight encryption \cite{bost2015machine} & \checkmark & -  & \checkmark \\
%  \hline
%  \multirow{3}{*}{Training} & Distributed training \cite{hamm2015crowd, shokri2015privacy, he2019model} & \checkmark & - & \checkmark \\
%  & Secure enclaves \cite{hunt2018chiron, ohrimenko2016oblivious} & - & \checkmark  & \checkmark \\
%  & Differential privacy perturbation \cite{abadi2016deep}  & \checkmark & \checkmark & - \\
%  %& Noise induction \cite{mireshghallah2020shredder} & - & \checkmark  & \checkmark \\
%  \hline
%  \multirow{2}{*}{Inference} & Output perturbation \cite{wu2016methodology, jia2019memguard, yang2020defending} & - & \checkmark  & \checkmark \\
%  & \textbf{\pelican (this paper)} & \checkmark & \checkmark & \checkmark \\
% \bottomrule
% \end{tabular}
% \vspace{-2mm}
% \label{tab:prior_defenses}
% \end{table}

Prior defenses against model inversion attacks have been limited and problem specific \cite{reza2019privacy, he2019model, yang2020defending, zhao2020tradeoff}. Zhao et al. presented a general attribute obfuscation framework using adversarial representation learning to protect sensitive attributes \cite{zhao2020tradeoff}. Yang et al. recently proposed an autoencoder-based prediction purification system to defend against model inversion attacks by minimizing the dispersion in output confidence scores \cite{yang2020defending}. The purifier is trained by minimizing the inversion attack accuracy and does not coincide with model training. Other defenses that have been proposed to prevent membership inference attacks may be relevant to model inversion attacks as well. We summarize these in Table \ref{tab:prior_defenses}.

Existing defense methods that require changes to the data, such as data obfuscation \cite{zhang2018privacy, zhao2020tradeoff} or encryption \cite{bost2015machine}, do not apply in this application since the output needs to be accessible to the honest-but-curious service provider. Additionally, existing differential privacy-based solutions only apply to multi-user models. In this work, we focus on post-hoc privacy preserving methods that are independent of the trained personalized models. Prior work in this domain \cite{jia2019memguard, yang2020defending} induce additional complexity of training noise induction models and are less feasible in applications where the model is on a resource-constrained mobile device. 

%Section 7
\section{Conclusion}
In this work, we examined the privacy implications of personalized models in distributed mobile services by proposing time-series based model inversion attacks. Our results demonstrated that such attacks can be used to recover historical mobility patterns that may be considered private by the user. We proposed a distributed framework, \pelican, that learns and deploys transfer learning-based personalized ML models in a privacy preserving manner on resource-constrained mobile devices. In \pelican, we introduced a novel privacy enhancement to thwart model inversion attacks. Our evaluation of \pelican using real world traces for location-aware mobile services showed that \pelican reduces privacy leakage substantially. 

% We would also like to acknowledge the higher efficacy of inference-based attacks in mobility applications. Mobility data is skewed \cite{trivedi2020empirical} and there is evidence suggesting the deep models memorize specific instances from the training data, typically outliers from the main distribution \cite{arpit2017closer}. Developing more generalizable personalized models in future work will lead to lower privacy implications. 

\section*{Acknowledgment}
We thank the anonymous reviewers for helpful comments. We also thank Priyanka Mammen, Amee Trivedi, Meet Vadera and Walid Hanafy for their feedback. This research was supported in part by NSF grants 1836752, 1763834, Army Research Lab contract W911NF-17-2-0196 and the United States Air Force under contract no. FA8750-17-C-0120.  Any opinions, findings and conclusions or recommendations expressed in this material are those of the authors and do not necessarily reflect the views of the funding agencies.
\balance
\typeout{}
\bibliographystyle{IEEEtran}
\bibliography{ref}

\begin{thebibliography}{10}
\providecommand{\url}[1]{#1}
\csname url@samestyle\endcsname
\providecommand{\newblock}{\relax}
\providecommand{\bibinfo}[2]{#2}
\providecommand{\BIBentrySTDinterwordspacing}{\spaceskip=0pt\relax}
\providecommand{\BIBentryALTinterwordstretchfactor}{4}
\providecommand{\BIBentryALTinterwordspacing}{\spaceskip=\fontdimen2\font plus
\BIBentryALTinterwordstretchfactor\fontdimen3\font minus
  \fontdimen4\font\relax}
\providecommand{\BIBforeignlanguage}[2]{{%
\expandafter\ifx\csname l@#1\endcsname\relax
\typeout{** WARNING: IEEEtran.bst: No hyphenation pattern has been}%
\typeout{** loaded for the language `#1'. Using the pattern for}%
\typeout{** the default language instead.}%
\else
\language=\csname l@#1\endcsname
\fi
#2}}
\providecommand{\BIBdecl}{\relax}
\BIBdecl

\bibitem{zhao2020go}
P.~Zhao, A.~Luo, Y.~Liu, F.~Zhuang, J.~Xu, Z.~Li, V.~S. Sheng, and X.~Zhou,
  ``Where to go next: A spatio-temporal gated network for next poi
  recommendation,'' \emph{IEEE Transactions on Knowledge and Data Engineering},
  2020.

\bibitem{dai2019machine}
X.~Dai, I.~Spasi{\'c}, B.~Meyer, S.~Chapman, and F.~Andres, ``Machine learning
  on mobile: An on-device inference app for skin cancer detection,'' in
  \emph{IEEE International Conference on Fog and Mobile Edge Computing}, 2019.

\bibitem{yoon2017efficient}
S.~Yoon, H.~Yun, Y.~Kim, G.-t. Park, and K.~Jung, ``Efficient transfer learning
  schemes for personalized language modeling using recurrent neural network,''
  \emph{AAAI Workshop on Crowdsourcing, Deep Learning, and Artificial
  Intelligence Agents}, 2017.

\bibitem{fredrikson2015model}
M.~Fredrikson, S.~Jha, and T.~Ristenpart, ``Model inversion attacks that
  exploit confidence information and basic countermeasures,'' in \emph{ACM
  SIGSAC Conference on Computer and Communications Security}, 2015.

\bibitem{gambs2012next}
S.~Gambs, M.-O. Killijian, and M.~N. del Prado~Cortez, ``Next place prediction
  using mobility {M}arkov chains,'' in \emph{First Workshop on Measurement,
  Privacy, and Mobility}, 2012.

\bibitem{mathew2012predicting}
W.~Mathew, R.~Raposo, and B.~Martins, ``Predicting future locations with hidden
  {M}arkov models,'' in \emph{ACM Conference on Ubiquitous Computing}, 2012.

\bibitem{song2016deeptransport}
X.~Song, H.~Kanasugi, and R.~Shibasaki, ``Deeptransport: Prediction and
  simulation of human mobility and transportation mode at a citywide level,''
  in \emph{International Joint Conference on Artificial Intelligence}, 2016.

\bibitem{hochreiter1997long}
S.~Hochreiter and J.~Schmidhuber, ``Long short-term memory,'' \emph{Neural
  Computation}, 1997.

\bibitem{kong2018hst}
D.~Kong and F.~Wu, ``{HST-LSTM}: A hierarchical spatial-temporal long-short
  term memory network for location prediction,'' in \emph{International Joint
  Conferences on Artificial Intelligence}, 2018.

\bibitem{trivedi2020empirical}
A.~Trivedi, J.~Gummeson, and P.~Shenoy, ``Empirical characterization of
  mobility of multi-device internet users,'' \emph{arXiv preprint
  arXiv:2003.08512}, 2020.

\bibitem{feng2020pmf}
J.~Feng, C.~Rong, F.~Sun, D.~Guo, and Y.~Li, ``{PMF}: A privacy-preserving
  human mobility prediction framework via federated learning,'' \emph{ACM on
  Interactive, Mobile, Wearable and Ubiquitous Technologies}, 2020.

\bibitem{vallon2017machine}
C.~Vallon, Z.~Ercan, A.~Carvalho, and F.~Borrelli, ``A machine learning
  approach for personalized autonomous lane change initiation and control,'' in
  \emph{IEEE Intelligent Vehicles Symposium}, 2017.

\bibitem{rudovic2018personalized}
O.~Rudovic, J.~Lee, M.~Dai, B.~Schuller, and R.~W. Picard, ``Personalized
  machine learning for robot perception of affect and engagement in autism
  therapy,'' \emph{Science Robotics}, 2018.

\bibitem{sarker2019effectiveness}
I.~H. Sarker, A.~Kayes, and P.~Watters, ``Effectiveness analysis of machine
  learning classification models for predicting personalized context-aware
  smartphone usage,'' \emph{Journal of Big Data}, 2019.

\bibitem{shokri2017membership}
R.~Shokri, M.~Stronati, C.~Song, and V.~Shmatikov, ``Membership inference
  attacks against machine learning models,'' in \emph{IEEE Symposium on
  Security and Privacy}, 2017.

\bibitem{fredrikson2014privacy}
M.~Fredrikson, E.~Lantz, S.~Jha, S.~Lin, D.~Page, and T.~Ristenpart, ``Privacy
  in pharmacogenetics: An end-to-end case study of personalized warfarin
  dosing,'' in \emph{{USENIX} Security Symposium}, 2014.

\bibitem{wu2016methodology}
X.~Wu, M.~Fredrikson, S.~Jha, and J.~F. Naughton, ``A methodology for
  formalizing model-inversion attacks,'' in \emph{IEEE Computer Security
  Foundations Symposium}, 2016.

\bibitem{hidano2017model}
S.~Hidano, T.~Murakami, S.~Katsumata, S.~Kiyomoto, and G.~Hanaoka, ``Model
  inversion attacks for prediction systems: Without knowledge of non-sensitive
  attributes,'' in \emph{International Conference on Privacy, Security and
  Trust}, 2017.

\bibitem{yang2019neural}
Z.~Yang, J.~Zhang, E.-C. Chang, and Z.~Liang, ``Neural network inversion in
  adversarial setting via background knowledge alignment,'' in \emph{ACM SIGSAC
  Conference on Computer and Communications Security}, 2019.

\bibitem{he2019model}
Z.~He, T.~Zhang, and R.~B. Lee, ``Model inversion attacks against collaborative
  inference,'' in \emph{Computer Security Applications Conference}, 2019.

\bibitem{zhang2020secret}
Y.~Zhang, R.~Jia, H.~Pei, W.~Wang, B.~Li, and D.~Song, ``The secret revealer:
  generative model-inversion attacks against deep neural networks,'' in
  \emph{IEEE/CVF Conference on Computer Vision and Pattern Recognition}, 2020.

\bibitem{zhang2018deeptravel}
H.~Zhang, H.~Wu, W.~Sun, and B.~Zheng, ``{DEEPTRAVEL}: a neural network based
  travel time estimation model with auxiliary supervision,''
  \emph{International Joint Conference on Artificial Intelligence}, 2018.

\bibitem{tan2018survey}
C.~Tan, F.~Sun, T.~Kong, W.~Zhang, C.~Yang, and C.~Liu, ``A survey on deep
  transfer learning,'' in \emph{International Conference on Artificial Neural
  Networks}, 2018.

\bibitem{platt1999probabilistic}
J.~Platt \emph{et~al.}, ``Probabilistic outputs for support vector machines and
  comparisons to regularized likelihood methods,'' \emph{Advances in Large
  Margin Classifiers}, 1999.

\bibitem{triastcyn2018generating}
A.~Triastcyn and B.~Faltings, ``Generating artificial data for private deep
  learning,'' \emph{PAL: Privacy-Enhancing Artificial Intelligence and Language
  Technologies}, 2019.

\bibitem{zhang2018differentially}
X.~Zhang, S.~Ji, and T.~Wang, ``Differentially private releasing via deep
  generative model (technical report),'' \emph{arXiv preprint
  arXiv:1801.01594}, 2018.

\bibitem{reza2019privacy}
K.~J. Reza, M.~Z. Islam, and V.~Estivill-Castro, ``Privacy preservation of
  social network users against attribute inference attacks via malicious data
  mining,'' in \emph{International Conference on Information Systems Security
  and Privacy (ICISSP)}, 2019.

\bibitem{zhao2020tradeoff}
H.~Zhao, J.~Chi, Y.~Tian, and G.~J. Gordon, ``Trade-offs and guarantees of
  adversarial representation learning for information obfuscation,''
  \emph{Neural Information Processing Systems (NeurIPS)}, 2020.

\bibitem{zhang2018privacy}
T.~Zhang, Z.~He, and R.~B. Lee, ``Privacy-preserving machine learning through
  data obfuscation,'' \emph{arXiv preprint arXiv:1807.01860}, 2018.

\bibitem{bost2015machine}
R.~Bost, R.~A. Popa, S.~Tu, and S.~Goldwasser, ``Machine learning
  classification over encrypted data,'' in \emph{Network and Distributed System
  Security Symposium}, 2015.

\bibitem{hamm2015crowd}
J.~Hamm, A.~C. Champion, G.~Chen, M.~Belkin, and D.~Xuan, ``Crowd-{ML}: A
  privacy-preserving learning framework for a crowd of smart devices,'' in
  \emph{International Conference on Distributed Computing Systems}, 2015.

\bibitem{shokri2015privacy}
R.~Shokri and V.~Shmatikov, ``Privacy-preserving deep learning,'' in \emph{ACM
  SIGSAC Conference on Computer and Communications Security}, 2015.

\bibitem{hunt2018chiron}
T.~Hunt, C.~Song, R.~Shokri, V.~Shmatikov, and E.~Witchel, ``Chiron:
  Privacy-preserving machine learning as a service,'' \emph{arXiv preprint
  arXiv:1803.05961}, 2018.

\bibitem{ohrimenko2016oblivious}
O.~Ohrimenko, F.~Schuster, C.~Fournet, A.~Mehta, S.~Nowozin, K.~Vaswani, and
  M.~Costa, ``Oblivious multi-party machine learning on trusted processors,''
  in \emph{{USENIX} Security Symposium}, 2016.

\bibitem{abadi2016deep}
M.~Abadi, A.~Chu, I.~Goodfellow, H.~B. McMahan, I.~Mironov, K.~Talwar, and
  L.~Zhang, ``Deep learning with differential privacy,'' in \emph{ACM SIGSAC
  Conference on Computer and Communications Security}, 2016.

\bibitem{jia2019memguard}
J.~Jia, A.~Salem, M.~Backes, Y.~Zhang, and N.~Z. Gong, ``Memguard: Defending
  against black-box membership inference attacks via adversarial examples,'' in
  \emph{ACM SIGSAC Conference on Computer and Communications Security}, 2019.

\bibitem{yang2020defending}
Z.~Yang, B.~Shao, B.~Xuan, E.-C. Chang, and F.~Zhang, ``Defending model
  inversion and membership inference attacks via prediction purification,''
  \emph{arXiv preprint arXiv:2005.03915}, 2020.

\end{thebibliography}

\end{document}